\documentclass[twocolumn,showpacs,preprintnumbers,amsmath,amssymb]{revtex4}

\usepackage{graphicx}
\usepackage{dcolumn}
\usepackage{bm}

\begin{document}

\title{Quantum correlations for arbitrarily high-dimensional Bell inequality}
\author{Che-Ming Li$^{1}$}
\author{Der-San Chuu$^{1}$}
\author{Yueh-Nan Chen$^{2}$}
\affiliation{$^{1}$Department of Electrophysics, National Chiao Tung University, Hsinchu
30050, Taiwan}
\affiliation{$^{2}$Department of Physics and National Center for Theoretical Sciences,
National Cheng Kung University, Tainan 701, Taiwan}
\affiliation{Department of Electrophysics, National Chiao Tung University, Hsinchu 30050,
Taiwan}
\date{\today }

\begin{abstract}
We analyze the correlation structure of bipartite arbitrary-dimensional Bell
inequalities via novel conditions of correlations in terms of differences of
joint probabilities called \textit{correlators}. The conditions of
correlations are shown to be necessary for the multi-level Bell state. In
particular, we find that the bipartite arbitrary-dimensional Bell-type
inequalities introduced by Collins-Gisin-Linden-Massar-Popescu [Phys. Rev.
Lett. \textbf{88}, 040404 (2002)] and Son-Lee-Kim [Phys. Rev. Lett. \textbf{
96}, 060406 (2006)] are composed of correlators, and we reveal that the
maximal violations by the Bell state just fulfill the conditions of quantum
correlations. Correlators can be considered as essential elements of Bell
inequalities.
\end{abstract}

\maketitle

\section{Introduction}

The remarkable properties of entanglement goes essentially beyond the
classical correlation constrained by two plausible assumptions, namely 
\textit{locality} and \textit{realism} (local realism) \cite{bell}. Local
realism is also the central view of Einstein, Podolsky, and Rosen (EPR) \cite
{epr} on the quantum mechanics. The assumption of \textit{realism} states
that outcomes of measurements are predeterministic, and the one of locality
says that a measurement performed by one party of a system does not
influence the result of the measurement performed by another party. By the
assumptions of local realism, the \textit{Bell inequalities} \cite
{bell,chsh,bi} for two-level systems have been proposed to
experimentally invalidate the point of view of EPR and to show that quantum
mechanics is \textit{not} locally realistic. For the aspect of quantum
information processing \cite{qip}, the nonlocal features of quantum
correlations enable people to perform high-security and novel quantum
communication \cite{ek,tel}. Moreover, it helps to solve the problems that
have no solutions in classical information theory \cite{cab}.

In addition to entanglement for quantum two-level systems (qubits), entangled
quantum multi-level systems (qudits) attract much attention for their
nonlocal characters \cite{collins,son,dghz} and advantages in quantum
information processing \cite{qdexp}. Collins \textit{et al.} \cite{collins}
have reformulate Bell inequalities to construct a large family of
multi-level inequalities in terms of a novel constraint for local-realistic
theories called Collins-Gisin-Linden-Massar-Popescu (CGLMP) inequality.
Recently, Son, Lee, and Kim (SLK) \cite{son} presented generic Bell
inequalities and their variants for arbitrary high-dimensional systems
through the generalized GHZ nonlocality \cite{gghz}.

In this work, we adopt a different approach to multilevel Bell inequalities.
We address the following question. What are the essential properties of
quantum correlations of entangled qudits that can be defined concretely and
be detected efficiently. We wonder whether these essential features can be
detected by Bell inequalities, i.e., whether the kernels of Bell
inequalities consist some correlation conditions that are necessary for the
entangled qudits. In order to attain this aim, we use novel conditions of
correlations in terms of pairs of difference of joint probabilities called 
\textit{correlators} to investigate correlation structure of multi-level
Bell state and Bell inequalities. These correlators can be measured locally
by each party, and by which the dependent properties of qudits can be
revealed in different directions of measurements. In
particular, we show that the CGLMP and SLK inequalities are comprised of
conditions of quantum correlations in terms of correlators. In the
following, an introduction to the conditions of correlations will be given
as a preliminary to further discussions and results.

\section{Correlation condition}

Before proceeding further, let us revisit the scenario of a two-party
Bell-type experiment for identifying the correlations between outcomes of
measurements. Therein, measurements on each spatially-separated particle are
assumed to be performed with two distinct results from different
observables. In each run of the experiment, the first observer chooses $V_{1}
$ and the second one chooses $V_{2}$ for their local measurements on their
particles respectively. After measurements, a set of results $v_{1}$ and $
v_{2}$, which can be either $0$ or $1$, is acquired. If sufficient runs of
such measurements have been made under the chosen local measurement setting,
the correlation between experimental outcomes can be revealed through the
analytical analysis of experimental records. In analogy, the multi-level
Bell type experiments work in the same way as mentioned above. The key idea
of this work is to utilize the correlation relations in terms of differences
of joint probabilities, i.e. correlators. We define two correlators in terms
of the differences of joint probabilities, which are given by 
\begin{equation}
C_{0}=P(0,0)-P(1,0)\;\text{and}\;C_{1}=P(1,1)-P(0,1),
\end{equation}
where $P(v_{1},v_{2})$ denote the joint probabilities for obtaining the sets
of results $(v_{1},v_{2})$ under a given local measurement setting. We can
show that outcomes of measurements performed on a system composed of \emph{\
two uncorrelated parts} must satisfy the following criteria: 
\begin{equation}
C_{0}\geq0\;\text{and}\;C_{1}\leq0\:\:\:\text{or}\:\:\:C_{0}\leq0\;\text{and}
\;C_{1}\geq0.
\end{equation}
To prove it, note that, for two uncorrelated parts, $C_{0}$ and $C_{1}$ can
be recast as: 
\begin{equation}
C_{0}=[P_{1}(0)-P_{1}(1)]P_{2}(0), C_{1}=[P_{1}(1)-P_{1}(0)]P_{2}(1), 
\nonumber
\end{equation}
where $P_{k}(v_{k})$ represent the probabilities to get a specific
measurement result for party $k=1,2$. Since $P_{k}(v_{k})\geq0$, thus we
always get $C_{0}C_{1}\leq0$, hence it ends the proof. In analogy, we can
formulate another set of correlators which is dual to the former one by $
\bar{C_{0}}=P(0,0)-P(0,1)$ and $\bar{C_{1}}=P(1,1)-P(1,0)$. Using these
equations, we also can set a condition for correlation between measurements
on pairs.

It is clear that, if the value of the product, $C_{0}C_{1}$, is positive,
one can assert that there must be correlations between the outcomes of the
measurements in the composite system in some way. In the quantum regime, we
consider $C_{0}$ and $C_{1}$ for a two-qubit pure entangled state:
\begin{equation}
\left|
\psi \right\rangle =\sin (\xi )\left| 0_{1}0_{2}\right\rangle_{z} +\cos (\xi)\left| 1_{1}1_{2}\right\rangle_{z},\nonumber
\end{equation}
where $\left|v_{1}v_{2}\right\rangle_{z}=\left|v_{1}\right\rangle_{1z}\otimes\left|v_{2}
\right\rangle_{2z}$ and $\left|v_{k}\right\rangle_{kz} $ is the eigenstate
of Pauli-operator $\sigma _{z}$ with eigenvalue $(-1)^{v_{k}}$ for the $k^{
\text{th}}$ party. If the local measurement setting is chosen as $(\sigma
_{x}$,$\sigma _{x})$, we obtain a violation of the criterion (2) by
\begin{equation}
C_{0}=C_{1}=\sin (2\xi )/2,\nonumber
\end{equation}
hence it turns out that the sum of $C_{0}$ and $
C_{1}$, denoted by $C^{(x)}$, equals to $\sin (2\xi )$. It is apparent that
the criteria can also be violated by
\begin{equation}
C_{0}=\sin ^{2}(\xi ),C_{1}=\cos^{2}(\xi),\nonumber
\end{equation}
and that $C^{(z)}=1$ under the setting $(\sigma _{z}$,$\sigma
_{z})$. Thus, we have, $C_{\psi }\equiv C^{(x)}+C^{(z)}=1+\sin (2\xi )$, for
the state $\left|\psi\right\rangle$. On the other hand, for example, the
corresponding $C^{(z)}$ for the \emph{separable} state $\rho_{\psi}=\sin
^{2}(\xi)\left| 0_{1}0_{2}\right\rangle_{zz}\left\langle 0_{1}0_{2}\right|
+\cos ^{2}(\xi)\left|
1_{1}1_{2}\right\rangle_{zz}\left\langle1_{1}1_{2}\right| $ also exhibits
the same correlation in the results under the setting $(\sigma _{z}$,$\sigma
_{z})$ and is also equal to one. However, since the corresponding $C_{0}$
and $C_{1}$ are both zero under the local setting $(\sigma _{x}$,$\sigma
_{x})$, the correlation is erased. Hence the summation of $C^{(x)}$ and $
C^{(z)}$, denoted by $C_{\rho }$, is equal to $1$. With the fact that $C_{\psi }>C_{\rho_{\psi} }$, we then can distinguish the pure entangled
state $\psi $ from the mixed $\rho_{\psi}$.

The above scenario for telling entanglement involves the summation of
criteria, $C_{0}$ and $C_{1}$ under two measurement settings. Since
entanglement manifests itself via quantum correlation in different
directions of measurements, it makes $C_{\psi }>C_{\rho_{\psi}}$. For
general cases, as will be discussed, we can prepare more settings of local
measurements and introduce more terms with the same meanings as $C_{0}$ and $C_{1}$ to investigate quantum correlations embedded in entangled states and
perform identification of their nonlocal properties further.

\section{Quantum correlation of bipartite arbitrary-dimensional Bell state}

The necessary conditions constructed by Eq. (1) for a system composed of
two-level independent pair can be extended to general ones. We give a set of
correlators for a system composed of two $d$-level parts, 
\begin{eqnarray}
&&C_{m}^{(12)}=P(v_{1}^{(1)}\doteq -m,v_{2}^{(2)}=m)  \nonumber \\
&&\quad\quad\quad-P(v_{1}^{(1)}\doteq 1-m,v_{2}^{(2)}=m),
\end{eqnarray}
for $m=0,1,...,d-1$, where $\doteq$ denotes equality modulus $d$. The
superscripts, $(12)$, $(1)$, and $(2)$, mean that the first measurement, $
V_{1}^{(1)}$, and the second measurement, $V_{2}^{(2)}$, have been selected
from two choices of each party. For a system composed of two \emph{independent} $d$-level parts, it must \emph{not} satisfy the following
condition: either $C_{m}>0$ or $C_{m}<0$ for all $m$ 's. If either of the
above conditions is satisfied by a bipartite system, there must be
correlations between the measurement outcomes. Let us give a concrete
example with $d=3$ for above statement. From Eq. (3), we represent the
correlators explicitly by 
\begin{eqnarray}
&&C_{0}^{(12)}=P(0,0)-P(1,0),  \nonumber \\
&&C_{1}^{(12)}=P(2,1)-P(0,1),  \nonumber \\
&&C_{2}^{(12)}=P(1,2)-P(2,2),  \nonumber
\end{eqnarray}
and furthermore if the particles composed of the bipartite system are
independent we have the following relations between probabilities 
\begin{eqnarray}
&&C_{0}^{(12)}=[P_{1}(0)-P_{1}(1)]P_{2}(0),  \nonumber \\
&&C_{1}^{(12)}=[P_{1}(2)-P_{1}(0)]P_{2}(1),  \nonumber \\
&&C_{2}^{(12)}=[P_{1}(1)-P_{1}(2)]P_{2}(2).  \nonumber
\end{eqnarray}
If we have the results: $P_{1}(1)>P_{1}(2)$ and $P_{1}(2)>P_{1}(0)$, it
turns out that $P_{1}(0)<P_{1}(1)$ which means that it is impossible to have 
$C_{m}^{(12)}>0$ for all $m$'s. Thus we can prove the statement for
arbitrary $d$ by the same way proposed above.

Other types of correlators similar to Eq. (3) can be readily formulated: 
\begin{eqnarray}
&&C_{m}^{(21)}=P(v_{1}^{(2)}\doteq d-m-1,v_{2}^{(1)}=m)  \nonumber \\
&&\quad\quad\quad-P(v_{1}^{(2)}\doteq -m,v_{2}^{(1)}=m), \\
&&C_{m}^{(ii)}=P(v_{1}^{(i)}\doteq -m,v_{2}^{(i)}=m)  \nonumber \\
&&\quad\quad\quad-P(v_{1}^{(i)}\doteq d-m-1,v_{2}^{(i)}=m),
\end{eqnarray}
for $i=1,2$.

Now, through the derived correlators, let us progress towards analysis of
the correlation structure of the bipartite arbitrary-dimensional Bell state: 
\begin{equation}
\left\vert \psi _{d}\right\rangle =\frac{1}{\sqrt{d}}\sum_{v=0}^{d-1}\left
\vert v\right\rangle _{1z}\otimes \left\vert v\right\rangle _{2z}.
\end{equation}
We represent the state $\left\vert \psi _{d}\right\rangle $ in the following
eigenbasis of some observable $V_{k}^{(q)}$, 
\begin{equation}
\left\vert l\right\rangle _{kq}=\frac{1}{d}\sum_{m=0}^{d-1}\text{exp}[i\frac{
2\pi m}{d}(l+n_{k}^{(q)})]\left\vert m\right\rangle _{kz},
\end{equation}
and then the joint probabilities for obtaining the measured outcome $
(v_{1}^{(i)},v_{2}^{(j)})$ for the state $\left\vert \psi _{d}\right\rangle $
are given by \cite{collins} 
\begin{eqnarray}
&&P_{\psi _{d}}(v_{1}^{(i)},v_{2}^{(j)})  \nonumber \\
&=&\frac{1}{2d^{3}\sin ^{2}[\frac{\pi }{d}
(v_{1}^{(i)}+v_{2}^{(j)}+n_{1}^{(i)}+n_{2}^{(j)})]},
\end{eqnarray}
where $(n_{1}^{(i)},n_{2}^{(j)})$ denote the local parameters of the local
measurement settings $(V_{1}^{(i)},V_{2}^{(j)})$. For the set of local
parameters given by 
\begin{equation}
n_{1}^{(1)}=0,n_{2}^{(1)}=1/4,n_{1}^{(2)}=1/2,n_{2}^{(2)}=-1/4,
\end{equation}
$C_{m}^{(ij)}$ can be evaluated analytically, and we arrive at 
\begin{equation}
C_{m,\psi _{d}}^{(ij)}=\frac{1}{2d^{3}}[\csc ^{2}(\frac{\pi }{4d})-\csc ^{2}(
\frac{3\pi }{4d})],
\end{equation}
for $i,j=1,2$. Since $C_{m}^{(ij)}>0$ for all $m$'s with any finite value of 
$d$, we ensure that outcomes of measurements performed on the particles of
the state $\left\vert \psi _{d}\right\rangle $ are dependent under four
different local measurement settings. Thus, we can consider each set of the
condition $C_{m}^{(ij)}>0$ as a necessary one of the Bell state $\left\vert
\psi _{d}\right\rangle $, and hence the corresponding correlation structure
of $\left\vert \psi _{d}\right\rangle $ could be specified concretely and
analytically via the correlators.

Furthermore, to compare the correlation embedded in $\left|\psi
_{d}\right\rangle$ predicted by quantum mechanics with the one by
local-realistic theories, we could utilize the necessary conditions proposed
above to achieve this aim. First, we combine all of the correlators involved
in the necessary conditions of $\left|\psi _{d}\right\rangle$ and evaluate
the summation of all $C_{m}^{(ij)}$ 's, 
\begin{equation}
C_{d}=C^{(11)}+C^{(12)}+C^{(21)}+C^{(22)},
\end{equation}
where $C^{(ij)}=\sum_{m=0}^{d-1}C_{m}^{(ij)}$. Then we have 
\begin{equation}
C_{d,\psi _{d}}=\frac{2}{d^{2}}[\csc^{2}(\frac{\pi}{4d})-\csc^{2}(\frac{3\pi
}{4d})].
\end{equation}
One can find that $C_{d,\psi _{d}}$ is an increasing function of $d$. For
instance, if $d=3$, one has $C_{3,\psi _{3}}\simeq 2.87293$. In the limit of
large $d$, we obtain, $\lim_{d\rightarrow \infty }C_{d,\psi _{d}}=(16/3\pi
)^{2}\simeq 2.88202$.

We proceed to consider the maximum value of $C_{d}$ by local-realistic
theories. The following derivation is based on deterministic local models,
since any probabilistic model can be converted into a deterministic one. We
substitute a chosen set, $(v_{1}^{(1)},v_{2}^{(1)},v_{1}^{(2)},v_{2}^{(2)})$
, into $C^{(ij)}$, and $C_{d}$ turns into 
\begin{eqnarray}
&&C_{d,\text{LR}}  \nonumber \\
&=&\delta \lbrack (v_{1}^{(1)}+v_{2}^{(1)})\text{mod }d,0]-\delta \lbrack
-(v_{1}^{(1)}+v_{2}^{(1)})\text{mod }d,1]  \nonumber \\
&&+\delta \lbrack (v_{1}^{(1)}+v_{2}^{(2)})\text{mod }d,0]-\delta \lbrack
(v_{1}^{(1)}+v_{2}^{(2)})\text{mod }d,1]  \nonumber \\
&&+\delta \lbrack (v_{1}^{(2)}+v_{2}^{(2)})\text{mod }d,0]-\delta \lbrack
-(v_{1}^{(2)}+v_{2}^{(2)})\text{mod }d,1]  \nonumber \\
&&+\delta \lbrack -(v_{1}^{(2)}+v_{2}^{(1)})\text{mod }d,1]-\delta \lbrack
(v_{1}^{(2)}+v_{2}^{(1)})\text{mod }d,0],  \nonumber \\
&&
\end{eqnarray}
where $\delta \lbrack x,y]$ represents the Kronecker delta symbol. It is
apparent that there are three non-vanishing terms at most among the four
positive delta functions under some specific condition for $(v_{1}^{(1)},v_{2}^{(1)},v_{1}^{(2)},v_{2}^{(2)})$. We also know that there
must exist one non-vanishing negative delta function in $C_{d,\text{LR}}$
under the same condition. Therefore, in the regime governed by
local-realistic theories, the value of $C_{d,\text{LR}}$ is bounded by $2$,
i.e., $C_{d,\text{LR}}\leq 2$.

From the above discussions, we realize that $C_{d,\psi _{d}}>C_{d,\text{LR} }
$. Therefore, the quantum correlations are stronger than the ones predicted
by the local-realistic theories. With this fact, the derived equation $C_{d}$
can be utilized to tell quantum correlations from classical ones. For $d=2$, 
$C_{2,\psi _{2}}=2\sqrt{2}$ and the equation $C_{d}$ is the same as that in
the CHSH \cite{chsh} inequality. Moreover, the $C^{(ij)}$ terms are just the
expectation values of the outcome products which appear in the CHSH
inequality. Then, we can reinterpret the correlation functions as a
summation of all $C_{0}^{(ij)}$ and $C_{1}^{(ij)}$ which formulate
correlation criteria for measurements on pairs. This idea can be applied to
arbitrary high-dimensional systems and to construct new types of correlation
functions, $C^{(ij)}$. Although the values of maximal quantum violation are
slightly smaller than the ones derived by Collins \textit{et al.} \cite{collins} and Fu \cite{fu}, the total number of joint probabilities required
by each of the presented correlation functions $C^{(ij)}$ is only $2d$,
which is much smaller than that in Fu's general correlation function, which
is about $O( d^{2})$. It implies that the proposed \emph{correlation}
functions include the essential parts of quantum correlation of the state $
\left|\psi _{d}\right\rangle$.

Another feature of the sum of all correlators will be discussed here is its
robustness to noise. If the state $\left\vert \psi _{d}\right\rangle $
suffered from white noise and turns into a mixed one in the form 
\begin{equation}
\rho =p_{\text{noise}}/d^{2}\openone+(1-p_{\text{noise}})\left\vert \psi
_{d}\rangle \langle \psi _{d}\right\vert ,
\end{equation}
where $p$ describes the noise fraction, the value of $C_{d}$ for state $\rho 
$ becomes $C_{d,\rho }=(1-p_{\text{noise}})C_{d,\psi _{d}}$. If the
criterion, $C_{d,\rho }>2$, i.e., $p_{\text{noise}}<1-2/C_{d,\psi _{d}}$, is
imposed on the system, one ensures that the mixed state still exhibits
quantum correlations in outcomes of measurements. For instance, to maintain
the quantum correlation for the limit of large $d$, the system must have $p_{
\text{noise}}<0.30604$.

Through the work by Masanes about \textit{tightness} of Bell inequality from
a geometric point of view \cite{tight}, we have examined our Bell-type
inequality. The result shows that the inequality is non-tight, i.e., it is
not an optimal detector of non-local-realistic correlation. The detailed
proof and discussions are given in the appendix.

\section{Correlation structure of CGLMP inequality}

Let us introduce more correlators like $C_{m}^{(ij)}$ to describe the
quantum correlations of $\left|\psi_{d}\right\rangle$. The first four sets
of correlators could be the one with the following form: 
\begin{eqnarray}
&&C_{m0}^{(ii)}=P(v_{1}^{(i)}\doteq m,v_{2}^{(i)}=m)  \nonumber \\
&&\quad\quad\quad-P(v_{1}^{(i)}\doteq m-1,v_{2}^{(i)}=m),  \nonumber \\
&&C_{m0}^{(12)}=P(v_{1}^{(1)}\doteq m,v_{2}^{(2)}=m)  \nonumber \\
&&\quad\quad\quad-P(v_{1}^{(1)}\doteq m+1,v_{2}^{(2)}=m),  \nonumber \\
&&C_{m0}^{(21)}=P(v_{1}^{(2)}\doteq m-1,v_{2}^{(1)}=m)  \nonumber \\
&&\quad\quad\quad-P(v_{1}^{(2)}\doteq m,v_{2}^{(1)}=m),
\end{eqnarray}
for $m=0,1,...,d-1$ and $i=1,2$. For $\left|\psi_{d}\right\rangle$, the
values of $C_{m0}^{(ij)}$ can be evaluated analytically under the same
measurement settings as the previous ones, and we have the result 
\begin{equation}
C_{m0,\psi_{d}}^{(ij)}=\frac{1}{2d^{3}}[\csc^{2}(\frac{\pi}{4d})-\csc^{2}(\frac{3\pi}{4d})],
\end{equation}
which is the same with $C_{m,\psi_{d}}^{(ij)}$ and $C_{m0,\psi_{d}}^{(ij)}>0$
for all $m$'s with any finite $d$. Thus we know that the particles of the
pair are dependent on each other. The second four sets of correlators are
introduced by 
\begin{eqnarray}
&&C_{m1}^{(ii)}=P(v_{1}^{(i)}\doteq m+1,v_{2}^{(i)}=m)  \nonumber \\
&&\quad\quad\quad-P(v_{1}^{(i)}\doteq m-2,v_{2}^{(i)}=m),  \nonumber \\
&&C_{m1}^{(12)}=P(v_{1}^{(1)}\doteq m-1,v_{2}^{(2)}=m)  \nonumber \\
&&\quad\quad\quad-P(v_{1}^{(1)}\doteq m+2,v_{2}^{(2)}=m),  \nonumber \\
&&C_{m1}^{(21)}=P(v_{1}^{(2)}\doteq m-2,v_{2}^{(1)}=m)  \nonumber \\
&&\quad\quad\quad-P(v_{1}^{(2)}\doteq m+1,v_{2}^{(1)}=m),
\end{eqnarray}
and the corresponding expectation values for the state $\left|\psi_{d}\right\rangle$ are 
\begin{equation}
C_{m1,\psi_{d}}^{(ij)}=\frac{1}{2d^{3}}[\csc^{2}(\frac{5\pi}{4d})-\csc^{2}(\frac{7\pi}{4d})],
\end{equation}
and are strictly greater than zero for all $m$'s with any finite $d$. Then
we obtain another four sets of correlators which could be utilized to
describe the dependence of the entangled pair.

Furthermore, let us progress towards to general sets of correlators which
are formulated by 
\begin{eqnarray}
&&C_{mk}^{(ii)}=P(v_{1}^{(i)}\doteq m+k,v_{2}^{(i)}=m)  \nonumber \\
&&\quad\quad\quad-P(v_{1}^{(i)}\doteq m-k-1,v_{2}^{(i)}=m),  \nonumber \\
&&C_{mk}^{(12)}=P(v_{1}^{(1)}\doteq m-k,v_{2}^{(2)}=m)  \nonumber \\
&&\quad\quad\quad-P(v_{1}^{(1)}\doteq m+k+1,v_{2}^{(2)}=m),  \nonumber \\
&&C_{mk}^{(21)}=P(v_{1}^{(2)}\doteq m-k-1,v_{2}^{(1)}=m)  \nonumber \\
&&\quad\quad\quad-P(v_{1}^{(2)}\doteq m+k,v_{2}^{(1)}=m),
\end{eqnarray}
for $k=0,...,\lfloor d/2\rfloor$. We deduce that the particles composed of
the Bell state are indeed dependent from the positive expectation values of
correlators with the following general forms: 
\begin{equation}
C_{mk,\psi_{d}}^{(ij)}=\frac{1}{2d^{3}}\{\csc^{2}[\frac{(1+4k)\pi}{4d}
]-\csc^{2}[\frac{(3+4k)\pi}{4d}]\}.
\end{equation}
Thus we can feature the quantum correlations embedded in the bipartite $d$-level Bell state in the $4(\lfloor d/2\rfloor+1)$ sets of correlators. Thus
we could take a linear combination of all of these sets of correlators as a
mens of identification: 
\begin{equation}
\mathsf{C}_{d}=\sum_{k=0}^{\lfloor
d/2\rfloor}\sum_{i,j=1}^{2}\sum_{m=0}^{d-1}f(k)C_{mk}^{(ij)},
\end{equation}
where $f(k)$ denotes the coefficient of combination which is function of $k$.

If we let $f(k)$ be 
\begin{equation}
f(k)=1-\frac{2k}{d-1},
\end{equation}
the summation of all of the correlators $\mathsf{C}_{d}$ becomes the kernel
of the CGLMP inequality \cite{collins}: 
\begin{equation}
\mathsf{C}_{CGLMP}=\sum_{k=0}^{\lfloor d/2\rfloor}\sum_{i,j=1}^{2}(1-\frac{2k}{d-1})C_{k}^{(ij)},
\end{equation}
where $C_{k}^{(ij)}=\sum_{m=0}^{d-1}C_{mk}^{(ij)}$. The local realistic
constraint proposed by Collins \textit{et al.} \cite{collins} specifies that
the correlations exhibited by local realistic theories have to satisfy the
condition: 
\begin{equation}
\mathsf{C}_{CGLMP,\text{LR}}\leq2.
\end{equation}
On the other hand, by Eq. (20), quantum correlations of the Bell state will
give a violation of the CGLMP inequality for arbitrary high-dimensional
systems. Thus, through Eqs. (20) and (21) and the related discussions, we
realize that the CGLMP inequality is composed of correlators for
correlations and know that the corresponding violations for the Bell state
just fulfill the conditions of quantum correlations of the entangled pair.

\section{Correlation structure of SLK inequality}

From the discussions in the previous sections, we could realize that the
features of entanglement of the Bell state are described by sets of
correlators with positive expectation values. Hence, we could generalize the
formulations of correlators for describing quantum correlations by the
following specification. The entanglement of the bipartite $d$-level Bell
state is featured in the correlators under different measurement settings: 
\begin{eqnarray}
&&C_{m}^{(l)}(\alpha ,\beta )  \nonumber \\
&=&P(v_{1}\doteq m+\alpha ,v_{2}=m)-P(v_{1}\doteq m+\beta ,v_{2}=m), 
\nonumber \\
&&
\end{eqnarray}
for $m=0,...,d-1$, where $l=[i,j]$ stands for measurement setting and $\alpha $ and $\beta $ are real numbers, and, most importantly, the values of
correlators for $\left\vert \psi _{d}\right\rangle $ strictly fulfill the
criterion 
\begin{equation}
C_{m,\psi _{d}}^{(l)}(\alpha ,\beta )>0,\text{ for }m=0,...,d-1,
\end{equation}
or 
\begin{equation}
C_{m,\psi _{d}}^{(l)}(\alpha ,\beta )<0,\text{ for }m=0,...,d-1.
\end{equation}
To have a compact form,  it should be noted that we have omitted the
denotations of measurement setting $(i,j)$ from the measured outcomes $v_{k}$
's. A linear combination of these correlators is utilized to identify the
state $\left\vert \psi _{d}\right\rangle $: 
\begin{equation}
\tilde{\mathsf{C}}_{d}=\sum_{l}\sum_{\alpha ,\beta }f_{l}(\alpha ,\beta
)C^{(l)}(\alpha ,\beta )
\end{equation}
where $C^{(l)}(\alpha ,\beta )=\sum_{m=0}^{d-1}C_{m}^{(l)}(\alpha ,\beta )$
and the coefficient of combination, $f_{l}(\alpha ,\beta )$, depends on $
\alpha $, $\beta $, and $l$.

Let us give a concrete example to show above formulation by the following
sum of correlators: 
\begin{equation}
\sum_{l}\sum_{\alpha =0}^{d-1}f_{l}(\alpha )\sum_{m=0}^{d-1}P(v_{1}\doteq
m+\alpha ,v_{2}=m),
\end{equation}
where 
\begin{eqnarray}
&&f_{l}(\alpha )=\sin (2\alpha _{l}\pi )[\cot (\alpha _{l}\pi /d)-\cot
(\alpha _{l}\pi )]/4, \\
&&\alpha _{l}=\nu +\alpha +\nu _{l},
\end{eqnarray}
$\nu $, and $\nu _{l}$ are constants. It is worth to note that 
\begin{equation}
\sum_{\alpha =0}^{d-1}f_{l}(\alpha )=0,
\end{equation}
which indicates that the sum of positive $f_{l}$'s and negative ones is zero
and implies that one can always have the following relation: 
\begin{widetext}
\begin{equation}
\sum_{\alpha=0}^{d-1}f_{l}(\alpha)\sum_{m=0}^{d-1}P(v_{1}\doteq m+\alpha,v_{2}=m)=\sum_{\alpha,\beta}f_{l}(\alpha,\beta)\sum_{m=0}^{d-1}[P(v_{1}\doteq m+\alpha,v_{2}=m)-P(v_{1}\doteq m+\beta,v_{2}=m)].
\end{equation}
\end{widetext} If we choose the same measurement settings as the previous
ones, please refer to Eqs. (7) and (9), the values of $C_{m}^{(l)}(\alpha
,\beta )$ for $\left\vert \psi _{d}\right\rangle $ strictly satisfy the
criteria $(26)$ or $(27)$ by the facts that 
\begin{equation}
C_{m,\psi _{d}}^{(ij)}(\alpha ,\beta )=P(v_{1}^{(i)}=\alpha
,v_{2}^{(j)}=0)-P(v_{1}^{(i)}=\beta ,v_{2}^{(j)}=0)
\end{equation}
for all $m$'s. Thus we conclude that Eq. (28) is indeed composed of
correlators for entanglement of the Bell state.

To investigate the meaning of the sum of correlators further, we assign
values to the parameters by $\nu=1/4$, $\nu_{11}=0$, $\nu_{22}=0$, $
\nu_{21}=1/2$, and $\nu_{12}=1/2$. We have the kernel of the SKL inequality: 
\begin{equation}
\mathsf{C}_{SLK}=\sum_{i,j=1}^{2}\sum_{\alpha,\beta}\sum_{m=0}^{d-1}f_{ij}(
\alpha,\beta)C^{(ij)}_{m}(\alpha,\beta).
\end{equation}
By a direct calculation, we have 
\begin{equation}
\sum_{\alpha=0}^{d-1}f_{ij}(\alpha)\sum_{m=0}^{d-1}P_{\psi_{d}}(v_{1}^{(i)}
\doteq m+\alpha,v_{2}^{(j)}=m)=(d-1)/4,
\end{equation}
for the Bell state, the value of the SLK kernel is 
\begin{equation}
\mathsf{C}_{SLK,\psi_{d}}=d-1.
\end{equation}
Son \textit{et al.} \cite{son} have shown that local-realistic theories
predict the value of the kernel by 
\begin{equation}
\mathsf{C}_{SLK,\text{LR}}\leq \frac{1}{4}[3\cot(\frac{\pi}{4d})-\cot(\frac{
\pi}{3d})]-1,
\end{equation}
which is called the SLK inequality. Thus the SLK inequality can be violated
by the Bell state by a factor: 
\begin{equation}
\lim_{d\rightarrow \infty}\frac{\mathsf{C}_{SLK,\text{LR}}}{\mathsf{C}
_{SLK,\psi_{d}}}=\frac{8}{3\pi},
\end{equation}
for arbitrary-high dimension.

\section{Summary}

In this work, we have analyzed the structures of Bell inequalities for
bipartite multi-level systems by conditions of correlations in terms of
correlators. We start with an investigation into the correlation properties
of the multi-level Bell state, and then we give specifications of the
correlation structure in terms of correlators. Through these correlators for
the Bell state, we construct Bell inequalities with fewer analyses of
measured outcomes. We also show that the CGLMP \cite{collins} and SLK \cite{son} inequalities are composed of correlation conditions in terms of
correlators. From the quantum mechanical point of view, we reveal that
correlators are the essential elements of the Bell inequalities for
arbitrarily high-diemensional systems.

\textit{Note added.---}During preparation of our manuscript, we were aware of one related structure of Bell inequalities for \textit{d}-level bipartite systems \cite{lee}.

\appendix*

\section{Tightness of Bell inequalities}

Every tight Bell inequality fulfills the following conditions \cite{tight}:

\textit{Condition 1.} All the generators of the convex polytope must belong
either to the half-space or to the hyperplane.

\textit{Condition 2.} There must be $4d(d-1)$ linear independent generators
among the ones that belong to the hyperplane.

On the other hand, non-tight Bell inequalities satisfy only Condition 1. Then, we will
examine the proposed BI by these conditions for tightness.

Firstly, we discuss condition 1 for the inequality. Although the same result
has been shown in our paper, i.e., the derivation of the bound of the
propose Bell inequality, we follow the approach presented by Masanes \cite{tight} for
completness. The summation of all correlators of quantum correlation can be
written as: 
\begin{eqnarray}
&&C_{d}  \nonumber \\
&&=P(v_{1}^{(1)}+v_{2}^{(1)}\doteq0)-P(v_{1}^{(1)}+v_{2}^{(1)}\doteq-1) 
\nonumber \\
&&\ +P(v_{1}^{(1)}+v_{2}^{(2)}\doteq0)-P(v_{1}^{(1)}+v_{2}^{(2)}\doteq1) 
\nonumber \\
&&\ +P(v_{1}^{(2)}+v_{2}^{(2)}\doteq0)-P(v_{1}^{(2)}+v_{2}^{(2)}\doteq-1) 
\nonumber \\
&&\ +P(v_{1}^{(2)}+v_{2}^{(1)}\doteq-1)-P(v_{1}^{(2)}+v_{2}^{(1)}\doteq0),
\end{eqnarray}
To have an explicit form of $C_{d}$ for further discussion, we define the
following variables: 
\begin{eqnarray}
&&\chi_{11}=v_{1}^{(1)}+v_{2}^{(1)}+\dot{d}_{11},  \nonumber \\
&&\chi_{12}=-v_{1}^{(1)}-v_{2}^{(2)}+\dot{d}_{12},  \nonumber \\
&&\chi_{22}=v_{1}^{(2)}+v_{2}^{(2)}+\dot{d}_{22},  \nonumber \\
&&\chi_{21}=-v_{1}^{(2)}-v_{2}^{(1)}-1+\dot{d}_{21},
\end{eqnarray}
where $\dot{d}_{ij}$ denotes a multiple of $d$ and $\chi_{ij}\in\{-1,0\}$
for $i,j=1,2$. In particular, the sum of the variables satisfies the
constrain: 
\begin{equation}
\sum_{i,j=1}^{2}\chi_{ij}\doteq-1.
\end{equation}
With the defined variables, $C_{d}$ is written as 
\begin{eqnarray}
&&C_{d}=\sum_{ij=1}^{2}P(\chi_{ij}=0)-P(\chi_{ij}=-1).
\end{eqnarray}
Next, we proceed to consider the extreme values of $C_{d}$ under the local
realistic theories. The all possible sets of $(\chi_{11},\chi_{12},
\chi_{22},\chi_{21})$ which fulfill the constraint of the sum of the
variables are as the following:

(i) three of the variables are $0$ and the rest is $-1$;

(ii) all of the variables are $-1$.\newline
The first class can be applied to arbitrary $d$, and, however, the second
one is only applicable for $d=3$. Thus, we have $C_{d,\text{LHV}}=2$ for the
class (i) and $C_{3, \text{LHV}}=-4$ for (ii), which mean that for all the
generators of the convex polytope for $C_{d,\text{LHV}}$ the value $C_{d,\text{LHV}}$ is equal or less than $2$. Thus the proposed Bell inequality fulfills the first condition.

Second, we consider the second condition for the Bell inequality. All the generators of the convex
polytope are written as 
\begin{equation}
\mathbf{G}=\left|v_{1}^{(1)},v_{2}^{(1)}\right\rangle\oplus
\left|v_{1}^{(1)},v_{2}^{(2)}\right\rangle\oplus
\left|v_{1}^{(2)},v_{2}^{(1)}\right\rangle\oplus
\left|v_{1}^{(2)},v_{2}^{(2)}\right\rangle,
\end{equation}
which, with the defined variables, can also be represented as the following: 
\begin{eqnarray}
&&\left|v_{1}^{(1)},\chi_{11}-v_{1}^{(1)}\right\rangle\oplus
\left|v_{1}^{(1)},-\chi_{12}-v_{1}^{(1)}\right\rangle  \nonumber \\
&&\
\oplus\left|v_{1}^{(1)}-\chi_{11}-\chi_{21}-1,\chi_{11}-v_{1}^{(1)}\right
\rangle  \nonumber \\
&&\ \
\oplus\left|v_{1}^{(1)}+\chi_{12}+\chi_{22},-\chi_{12}-v_{1}^{(1)}\right
\rangle,
\end{eqnarray}
where $\left|v_{1}^{(i)},v_{2}^{(j)}\right\rangle=\left|v_{1}^{(i)}\text{mod}
\ d\right\rangle\otimes\left|v_{2}^{(j)}\text{mod}\ d\right\rangle$. Through
a linear transformation with the preservation of orthogonality which is
given by 
\begin{eqnarray}
&&\sum_{v,k=0}^{d-1}(\left|v,\chi_{11}\right\rangle\left\langle
v,\chi_{11}-v\right|\oplus\left|v,\chi_{12}\right\rangle\left\langle
v,-\chi_{12}-v\right|  \nonumber \\
&&\ \ \ \oplus\left|v-\chi_{11},\chi_{21}\right\rangle\left\langle
v-\chi_{11}-\chi_{21}-1,\chi_{11}-v\right|  \nonumber \\
&&\ \ \ \oplus\left|v+\chi_{12},\chi_{22}\right\rangle\left\langle
v+\chi_{12}+\chi_{22},-\chi_{12}-v\right|),
\end{eqnarray}
$\mathbf{G}$ can be transformed into 
\begin{eqnarray}
&&\left|v_{1}^{(1)},\chi_{11}\right\rangle\oplus\left|v_{1}^{(1)},-\chi_{12}
\right\rangle  \nonumber \\
&&\ \ \ \ \oplus\left|v_{1}^{(1)}-\chi_{11},\chi_{21}\right\rangle\oplus
\left|v_{1}^{(1)}+\chi_{12},\chi_{22}\right\rangle.
\end{eqnarray}
The generators which satisfy $C_{d,\text{LHV}}=2$ are the ones with
variables belonging to (i). Thus, the generators contained in the hyperplane
are shown as: 
\begin{eqnarray}
&&\left|v,-1\right\rangle\oplus\left|v,0\right\rangle\oplus\left|v+1,0\right
\rangle\oplus\left|v,0\right\rangle, \\
&&\left|v,0\right\rangle\oplus\left|v,-1\right\rangle\oplus\left|v,0\right
\rangle\oplus\left|v-1,0\right\rangle, \\
&&\left|v,0\right\rangle\oplus\left|v,0\right\rangle\oplus\left|v,-1\right
\rangle\oplus\left|v,0\right\rangle, \\
&&\left|v,0\right\rangle\oplus\left|v,0\right\rangle\oplus\left|v,0\right
\rangle\oplus\left|v,-1\right\rangle,
\end{eqnarray}
for $v\in\{0,1,...,d-1\}$. The total number of linear independent generators
is $4d$ which is smaller than $4d(d-1)$ involved in the condition of
tightness. Then the proposed Bell inequality is non-tight.\newline
\newline

\end{document}